\documentstyle[aps,prl,preprint]{revtex}
\input {epsf.sty}
\tightenlines
\begin{document}
\draft
\title{Exclusion Statistics in a two-dimensional trapped Bose gas}
\author{S. Viefers}
\address{Department of Applied Physics, 
    Chalmers University of Technology and G\"oteborg University \\
    SE-412 96 G\"oteborg,
    Sweden  }
\author{T. H. Hansson}
\address{Physics Department, Stockholm University\\ Stockholm Center 
for Physics, Astronomy and Biotechnology\\ Roslagstullsbacken 21, 
S-106 91, Stockholm, Sweden  }
\author{J. M. Leinaas}
\address{Department of Physics,  University of Oslo \\
P.O. Box 1048 Blindern, N-0316 Oslo, Norway }
\date{\today}
\maketitle
\begin{abstract}
We briefly explain the notion of  exclusion statistics and in 
particular discuss the concept of an ideal exclusion statistics gas. We then 
review a recent work where it is demonstrated that  
a {\em two-dimensional} Bose 
gas with repulsive delta function interactions obeys ideal exclusion 
statistics, with a fractional parameter related to the interaction 
strength. 
\end{abstract}

\vspace{-11pt}

\newcommand{\half}{\frac 1 2 }
\newcommand{\eg}{{\em e.g.} }
\newcommand{\ie}{{\em i.e.} }
\newcommand{\etc} {{\em etc.}}

\newcommand{\noi}{\noindent}
\newcommand{\etal}{{\em et al.\ }}
\newcommand{\cf}{{\em cf. }}

\newcommand{\dd}[2]{{\rmd{#1}\over\rmd{#2}}}
\newcommand{\pdd}[2]{{\partial{#1}\over\partial{#2}}}
\newcommand{\pa}[1]{\partial_{#1}}
\newcommand{\pref}[1]{(\ref{#1})}

\newcommand{\bea}{\begin{eqnarray}}
\newcommand{\eea}{\end{eqnarray}}
\newcommand{\e}{\varepsilon}
\newcommand{\D}{\partial}
\newcommand{\pt}{\tilde p}
\newcommand{\lt}{\lambda_T}

\newcommand{\ee}{\end{eqnarray}}


\newcommand {\be}[1]{
      \begin{eqnarray} \mbox{$\label{#1}$}  }

\section{Introduction}
Low-dimensional electron
systems have been extensively studied during the last decades, both 
due to their practical importance and because of their exotic quantum 
mechanical properties.
Prominent examples are the 2D heterojunctions that at low temperature, 
and in high magnetic fields, exhibit the quantum Hall 
effect\cite{prange1}, and carbon nanotubes, which are quasi 
1d and (for certain geometries) are expected to behave as Luttinger 
liquids\cite{fisher1}. 

At the same time there has  been a tremendous advance in the 
techniques for trapping and cooling atoms, which has opened new vistas 
for the study of quantum gases, and in particular it is now  possible 
to create quasi-two dimensional atomic Bose gases.

A very interesting property of certain low-dimensional systems is 
that the elementary excitations obey exotic quantum statistics.
The most celebrated example is the quantum Hall effect, where 
the Laughlin quasi particles are  {\em anyons}\cite{leinaas1}, \ie particles
whose quantum statistics ``interpolates'' between that of bosons and 
fermions in the sense that their many-body wave function picks up
a general phase $\exp{(i\theta)}$ (rather than just 1 or $-1$) 
under particle exchange. 

Some years ago, Haldane proposed another type of generalized quantum 
statistics, 
so-called {\em (fractional)
exclusion statistics} (FES)\cite{haldane1}.
Rather than modifying the many-body
exchange phases, 
he generalized the Pauli principle, \ie introduced new  rules for
occupying single-particle quantum states. 
The basic idea of FES is that adding a number of particles, $
\Delta N$, to a system, blocks
$\Delta d$ of the states available  for the next particle according to the
linear relation $\Delta d = -g\Delta N$.
This corresponds to a repulsion between the particles in phase space,
and  only very special types of interactions give rise to this type
of exclusion of single particle states. One of Haldane's original 
examples was  a one-dimensional spin chain; moreover, he showed that the 
Laughlin quasi particles (or generally anyons in a strong magnetic 
field, \ie confined to the lowest Landau level) 
also obey exclusion statistics with an exclusion parameter 
$g = \theta/\pi$. 
While anyons for topological reasons can exist only
in one or two dimensions, there is no such restriction 
on exclusion statistics particles. Nevertheless, until some time ago, 
all established examples of FES were either strictly 
one-dimensional\cite{isakov3,FESrealiz}, 
or, as anyons in the lowest Landau level,  
effectively one-dimensional\cite{veigy1,johnson1,hansson1}.

In a recent work\cite{PRL} we established that  within a certain 
(experimentally
feasible) parameter range  a (quasi-) {\em two-dimensional} Bose gas
with repulsive delta function interaction is in fact an ideal FES 
gas where the exclusion parameter is determined by the interaction 
strength.

\noindent
The model Hamiltonian, given by
\be{ham}
H = \sum_{i=1}^N\left(\frac {p_i^2} {2m} + V(\vec r_i) \right)
  +\frac {\pi\hbar^2} m g \sum_{i<j}^N\delta^2(\vec r_i - \vec r_j
),
\ee
 has previously been used to describe atoms in Bose condensation 
experiments using highly asymmetric traps
\cite{haugset1,assym}. The particular parametrization of the delta function
potential is chosen as to reproduce the
s-wave scattering phase-shift in three dimensions, and the dimensionless
coupling $g$ is given by
\be{cc}
g=2\sqrt{\frac 2 \pi}\  \frac a {l_z} \ ,
\ee
where $a$ is the (three dimensional) scattering length and $l_z$ the out of
plane extension of the asymmetric trap, which is harmonic in the transverse
direction with a frequency $\omega =\hbar/ml_z^2$
\cite{haugset1,bhaduri2}. 
We shall assume
that the temperature is sufficiently high above the transition temperature
that the only  relevant mean field is the density, $n$, and that the
fluctuations  are small enough to be ignored. We shall return to these issues
in the discussion (section \ref{Dsec}).

Our study was motivated by the previous
observation\cite{bhaduri1,bhaduri2}
that, in a Thomas-Fermi approximation, a
two dimensional Fermi or Bose gas with short range repulsive
interactions has the same energy and number
density as an ideal FES gas (treated in the same approximation).
To reach our conclusion, we analyzed the statistical mechanics of 
the delta function gas, both by using a  
mean field approximation for the energy levels, and by
directly relating the thermodynamics to the quantum 
mechanical scattering phase shifts.

In order to make this presentation reasonably self contained, 
section \ref{FESsec}
gives a brief introduction to exclusion statistics and in particular 
to ideal FES gases. Section \ref{MFTsec} contains the mean field argument for 
the equivalence of \pref{ham} to an ideal FES gas, and in
section \ref{QMsec} we show that this result is consistent
with a full quantum mechanical treatment of the model within
a certain parameter regime. Finally, we discuss our results
and point out some open questions.

\section{Exclusion statistics}  \label{FESsec}
As already mentioned, exclusion statistics with
statistics parameter $g$ is defined by the relation\cite{haldane1}
\be{fesdef}
\Delta d(N) = -g\Delta N
\ee
stating that $d(N)$, the dimension of the single-particle Hilbert space
when $N-1$ states are occupied,
decreases by $g$ whenever a particle is added to the system.
Obviously, $g=1$ corresponds to fermions (Pauli blocking), while
$g=0$ describes Bose statistics. In general, $g$ can be any rational
or integer number.
The corresponding statistical weight, i.e. the number of ways in
which a given number of particles can be distributed over a given 
number of states, is\cite{isakov1,wu1} 
\be{W}
W = \frac{\left[D-(g-1)(N-1)  \right]!}{N!\left[D-gN+g-1 \right]!}
\ee
where $D=d(1)$.
Eq.\pref{W} is seen to interpolate between the familiar expressions for
bosons ($g=0$) and fermions ($g=1$) in accordance with Eq.\pref{fesdef}.

An {\em ideal} FES gas is defined\cite{isakov1} by demanding that this
generalized exclusion principle applies locally in energy space 
\ie to a set of degenerate or almost degenerate  energy levels. 
In the case of the quantum Hall effect, $g$ is, for 
topological reasons always rational while in our case, being related 
to the interaction strength, it can take any real value. For non 
rational $g$, $W$ in \pref{W} is not integer, and has no 
combinatorical interpretation, but is rather used to {\em define} the 
concept of FES. 

In direct analogy with the case of bosons and fermions, the statistical 
distribution function is  derived from \pref{W}
by maximizing the non-equilibrium entropy 
$S \equiv -k_B \sum_k \ln W_k$, where the sum is over all energy 
levels $\epsilon_k$, subject to the constraints of keeping the total 
energy and number of particles fixed\cite{isakov1,wu1}.
The entropy is found to be
\be{ent0}
S &=& \sum_k D_k\left[  \left( 1-(g-1)n_k \right)\ln \left( 1-(g-1)n_k \right)
   - n_k \ln n_k  
   -  \left( 1-g n_k \right) \ln \left( 1-g n_k \right)  \right],
\ee
where $n_k \equiv N_k/D_k$, and the resulting statistical
distribution function (or occupancy
factor) can be expressed
as\cite{isakov1,wu1,isakov4} $n_k = (w_k + g)^{-1}$ where
\be{dis}
w_k^g \left( 1+w_k \right)^{1-g} = \exp(\beta(\epsilon_k - \mu)),
\ee
with $\beta=1/k_B T$. 
The special cases $g=0,1$ correspond to the familiar distribution
functions of ideal bosons and fermions, respectively. At zero temperature
$n_k$ is a step function, 
\be{gst}
n_K^0 = \frac{1}{g} \, \theta(\epsilon_F - \epsilon_k),
\ee
where $\epsilon_F$ is the Fermi energy. This means that in the ground 
state, there are on average $1/g$ particles in each original quantum state.
In section \ref{MFTsec}, we shall derive the expression 
\pref{ent0} for the entropy starting from the 2D delta function gas 
\pref{ham}. 

The thermodynamics of exclusion statistics has been studied in a
number of papers\cite{wu1,isakov4,nayak1,isakov5}. A result that is
particularly interesting for our purposes, is the virial expansion 
(\ie the expansion of the pressure in powers of the density) of an
ideal FES gas. As shown by Isakov et al.\cite{isakov4}, the equation
of state can be evaluated exactly in the special case $D=\sigma$
where $D$ is the number of dimensions and $\sigma$ the power of the
dispersion, $\varepsilon(p) \propto p^{\sigma}$. Obviously the case we
are interested in, a non-relativistic two dimensional gas with 
$\varepsilon(p) = p^2/2m$, falls into this class. 
Thus, for $D = \sigma = 2$, the virial expansion of an ideal
FES gas takes the form
\be{FESvir}
\frac{P}{k_B T} &=& n + \frac{1}{4} (2g-1) \lt^2 n^2
  + \frac{1}{\lt^2}\sum_{p=3}^\infty \frac { B_{p-1} }{p!} (\lt^{2}n)^{p} \\
             &=& \frac{P_{\rm Bose}}{k_B T} + \frac{1}{2}g\lt^2 n^2, 
\ee
where $\lambda_T = \sqrt{2\pi\hbar^2\beta/m}$ is the thermal
wavelength.
Here it is also worth
mentioning that for ideal quantum gases in 2D (bosons, fermions
or ``exclusons''), the pressure equals the energy density $\cal{E}$.
Note that all statistics dependence in \pref{FESvir}
lies in the second virial coefficient,
that interpolates linearly between the familiar results for 
ideal bosons and fermions in 2D\cite{sen1}. 
In section \ref{QMsec} we shall derive such a linear relation for the 
2D delta function gas.

\section{Mean field theory} \label{MFTsec}
Before going into a  statistical mechanical analysis
of our model, we  give a simple thermodynamic mean field 
argument for why we expect exclusion statistics.
For simplicity we consider the case
with a constant external potential $V$, so that the
density, $n$, is also  constant. The crucial observation is then
that in a mean field approximation, and for a fixed number of particles,
the interaction energy term in \pref{ham} just amounts to a constant
shift of the energy density,
\be{eos}
{\cal E} &=& {\cal E}_{\rm Free Bose} +  \frac{\pi\hbar^2}{m} g n^2
\ .
\ee
Expressing this in terms of the thermal wavelength, 
we see that this shift is identical to
that in the FES virial expansion Eq.\pref{FESvir} if we 
identify the interaction strength
$g$ with the exclusion statistics parameter. In other words, our
interacting Bose system \pref{ham} has the same (mean field) energy
density as an ideal FES gas with statistics $g$.

Let us now consider the statistical mechanics
of \pref{ham}, and assume that the potential $V$ is
slowly varying compared with the thermal wavelength, $\lambda_{T}$.
We  then divide
the system into cells of area $b^2$, where
$\lambda_T \ll b \ll |\vec\nabla V /V|$, and study the
statistical mechanics in each cell.
In a mean field approximation,  the one-body Hamiltonian
in the cell $\ell$  becomes
\be{oneh}
H_{\ell} = \frac {p^2} {2m} + V(\vec r_\ell)
       + \frac {2\pi\hbar^2} m g\, n(\vec r_\ell) \ ,
\ee
with the corresponding energy $\epsilon^{\ell}$,
where we  approximate the potential in the cell with the constant
$V(\vec r_{\ell})$ with $\vec r_\ell$  the position of the center of the
cell.
Also $n(\vec r_{\ell})=N_{\ell}/b^{2}$ is the mean number density in the
cell $\ell$, with $N_{\ell}\gg 1$  the corresponding average number of
particles.
\footnote{In going from \pref{ham} to  \pref{oneh} it is important to
correctly incorporate the effect of Bose statistics\cite{griffin}.
There is an extra factor of 2 in the interaction term in the Heisenberg
equation
for the quantum field operator  as compared to the similar looking
Gross-Pitaevskii equation for a classical Bose field. This also implies an
extra factor of two in the mean field one-body Hamiltonian \pref{oneh},
and here we differ from the treatment in\cite{bhaduri2}.}
We want to demonstrate that the number of available single-particle
states in a box $\ell$ decreases with the number of particles as
in Haldane's definition of FES, Eq.\pref{fesdef}. To this end, we
write the quantized momenta in the box as 
$\vec p = (2\pi\hbar/b) \vec l$ with $l_x$ and $l_y$ integers.
The total number of states $d_{\ell}$ below some energy $\epsilon^{\ell}$ 
corresponding to a momentum $p$, is $d_\ell =\pi l^2$. Using
\pref{oneh} to eliminate $p$, one finds
\be{nstat}
d_\ell &=& 
\frac {mb^2} {2\pi\hbar^2}  [\epsilon^{\ell} - V(r_\ell)] - gN_\ell \\
       &=& d_{\ell,{\rm Bose}} - gN_\ell  \, .
\ee
Indeed this result indicates exclusion statistics. 
Note that an important ingredient of the derivation of $d_\ell$
is that the dimension matches the power of the dispersion,
in this case $D = \sigma = 2$.

Eq.\pref{nstat} describes the dependence of the {\em total} 
number of states below some energy on the {\em total} number of
particles in the box.
But, as explained in section \ref{FESsec}, for an ideal FES gas,
the Haldane exclusion property must hold locally
in energy space, so the state counting result \pref{nstat} is
a necessary but not sufficient condition. The strategy to prove
that the box Hamiltonian \pref{oneh} indeed describes an ideal
FES gas, is to first map the interacting system onto a
non-interacting one in terms of {\em renormalized quantum numbers}
(pseudo-energies) which display the exclusion statistics property.
Then we can use standard counting arguments to arrive at the 
expression \pref{ent0} for the entropy of an ideal FES gas. 

The starting point is the total energy, $E_{\ell}$ which is not 
simply the sum of the one particle energies $\epsilon_{i}^{\ell}$ 
of \pref{oneh}, but given by
\be{toten}
E_{\ell} = \sum_{i}\left[\epsilon_{i}^{\ell}
- \frac {\pi\hbar^2} m g\, n(\vec r_\ell)\right] \, ,
\ee  where the last term
compensates for the double counting of the interaction energy.
Recalling from above that the number of states below a given
energy is proportional to the corresponding kinetic energy,
we can rewrite Eq.\pref{toten} as 
\be{micen}
E_\ell =\sum_{i=1}^{N_\ell} \left( V(\vec r_\ell)
       + \frac {2\pi\hbar^2} {mb^2} [k_i + \frac g 2 N_\ell ]
       \right) \, ,
\ee
where $k_i = d_{\ell,i}$ is the number of available single-particle
states below the energy of particle $i$, and we choose to order the
labelling of particles in a given microstate such that 
$0\le k_1 \le k_2 \cdots \le k_{N_\ell}$.
Although the energy levels (the integers $\{ k_i \}$) are not equally
spaced, we assume that the box is large enough for this effect to
be negligible.

Next introduce the quantities $\tilde k_i$ by
\be{psenlab}
\tilde k_i = k_i + g\sum_j \theta(\tilde k_j - \tilde k_i) \ ,
\ee
with $\theta(x)$ being the step function,
and rewrite the energy as a sum of "pseudo-energies",
$\tilde\epsilon_{i}^{\ell}$,
\be{psen}
E_\ell &=& \sum_{i=1}^{N_\ell} \tilde\epsilon^\ell_i \\
\tilde\epsilon^\ell_i &=& V(r_\ell) + \frac {2\pi\hbar^2} {mb^2} \tilde
k_i \, .  \nonumber
\ee
Note that, expressed in terms of these pseudo-energies, the system
is formally non-interacting. The interaction term $\sim gN_{\ell}$
has been absorbed in the renormalized quantum numbers $\tilde k_i$.
The exclusion properties of this system are now manifest since \pref{psenlab}
implies that
the pseudo-energies must satisfy,
\be{excl}
\tilde\epsilon^{\ell}_{i+1}\ge \tilde\epsilon^{\ell}_i 
                             + \frac{2\pi}{mb^2} \hbar^2 g.
\ee

As mentioned above, the relation \pref{nstat} holds  only  because
the kinetic energy and the number density scale as the same power of
the cell size $b$.
This is true for the present case of particles with quadratic
dispersion in two dimensions,
but also for particles in one dimension with linear dispersion, which
is the case for anyons in the
lowest Landau level, or equivalently, chiral particles on a circle with an
$N^2$ type interaction \cite{hansson1}. In fact, the two models studied in
this paper and in \cite{hansson1} can be exactly mapped onto each
other by identifying the $\tilde k_{i}$ in \pref{psenlab} with the
``pseudo-momenta'' introduced in \cite{hansson1}.

The  proof that the exclusion property \pref{excl} corresponds
to an ideal FES gas as
defined in \cite{isakov1}, amounts to showing that the (non-equilibrium) 
entropy of our system equals that of an ideal FES gas. This derivation  
is a straightforward modification of the
one given in \cite{isakov2} for a multispecies system in the
fermionic representation: Going to a continuum description with
``momenta'' $p_i$ and ``pseudo-momenta'' $\pt_i$ defined as
\be{ppt}
p_i = \left( \frac{2\pi\hbar}{b} \right)^2 k_i ; \ \ \ \ \
\pt_i = \left( \frac{2\pi\hbar}{b} \right)^2 {\tilde k}_i,
\ee
replacing the sums over $k$ and $\tilde k$ by integrals in the
usual way and denoting the corresponding particle densities
in momentum space by $\nu(p)$ and $\rho({\tilde p})$, respectively,
one finds the continuum version of Eq.\pref{psenlab},
\be{psenc}
\pt = p + g \int d\pt' \rho(\pt') \theta(\pt - \pt') \ .
\ee
Furthermore one has to demand conservation of the number of particles
when changing variables from $p$ to $\pt$, \ie
$\nu(p) dp = \rho(\pt) d\pt$. Combining this with Eq.\pref{psenc} gives
\be{nurho}
\nu(p) = \frac{\rho(\pt)}{1 - g\rho(\pt)}.
\ee
Inserting this into the standard expression for the bosonic
non-equilibrium entropy,
\be{ent1}
S = -k\left( \frac{b}{2\pi\hbar} \right)^2
      \int dp \left[ \nu\ln\nu - (1+\nu)\ln(1+\nu) \right] \ ,
\ee
exactly reproduces the continuum version of the FES entropy \pref{ent0}
\be{ent2}
S &=& -k \left( \frac{b}{2\pi\hbar} \right)^2
         \int d\pt \left[ \rho\ln\rho 
      -  \left( 1-(g-1)\rho \right) \ln \left( 1-(g-1)\rho \right)
      +  \left( 1-g\rho \right)\ln\left( 1-g\rho \right)
      \right]  \ ,
\ee
from which all thermodynamics follows.

\section{Beyond mean field theory: Quantum mechanics}  \label{QMsec}
So far our analysis was entirely in the
context of mean field approximations.
It is an interesting question
whether the  full quantum problem
of a two dimensional gas with a delta function interaction
also allows a description
in terms of exclusion statistics in some range of temperatures.
In this section we show that this is indeed the case, by
computing the correction to the second virial coefficient
due to the interaction, and showing that in a certain
parameter regime, the result is consistent with Eq.\pref{FESvir}.
To this end, we first calculate the quasi-2D
scattering solution with a delta function potential to find the
s-wave scattering phase shift. The corresponding shift in the pressure
of the gas can be found from the Beth-Uhlenbeck formula\cite{landau1}
which relates the second virial coefficient to the scattering
phase shifts.
Although the interaction naively does not involve
any dimensionful parameter, it is  known that a pure delta function
interaction gives rise to short distance singularities and requires a
renormalization of the interaction strength which introduces a
renormalization scale and thus breaks scale invariance\cite{2d-delta}.
The two-dimensional s-wave scattering phase shift $\delta_0$
is  given by\cite{petrov2000}
\be{ph}
\cot \delta_0 = -\frac 4 {\pi g} + \frac 2 \pi \ln
\left(\frac {p\, l_z^{eff}} \hbar \right) \ ,
\ee
where $p$ is the relative momentum in the two body scattering process, and
$g$ is related to the (three dimensional) scattering length by \pref{cc}.
Note that the phase shift does depend on the
momentum $p$  via the renormalization scale,
$l_z^{eff}$ which up to a numerical factor equals the transverse 
extent $l_z$ of the quasi 2d system.  
The Beth-Uhlenbeck formula then gives
the following shift in the pressure due to interactions\cite{giacconi1},
\be{bu}
\Delta P = -(n\lambda_T)^2 \,   \frac {2kT} \pi
                        \int_0^\infty dp\, e^{-\frac {p^2} {mkT} }
            \frac{d\delta_0} {dp} \ .
\ee
In general, the integral in \pref{bu} is a function of $l_z/\lambda_T$,
but in the relevant parameter range, 
\be{range}
a \ll l_z \ll \lambda_T , 
\ee
it is
approximately constant, as can be seen by the following argument:
First partially integrate and note that the phase shift vanishes at 
$p=0$. For small $g$, \ie $a\ll l_{z}$ we consider the $O(g)$ 
correction to the ideal Bose gas:
\be{bu2}
\frac{d\Delta P}{dg} = -\frac{4}{m\pi} (n\lambda_T)^2 \,
                        \int_0^\infty dp\,p e^{-\frac {p^2} {mkT} }
            \frac{d\delta_0} {dg} \ .
\ee
where 
\be{ph2}
\frac{d\delta_0} {dg} = -\frac{4}{\pi g^{2}} \left[ 1 + 
\left( -\frac 4 {\pi g} + \frac 2 \pi \ln( p l_z^{eff}/ \hbar)\right)^{2}
\right]^{-1}.
\ee
Next note that the  integral has a high momentum cutoff at $p\sim 
\frac{\hbar}{\lambda_{T}}$ due to the exponential. 
This cutoff suppresses contributions from values of $p$ large enough
for the logarithm to give a sizable correction, since we have
to demand $l_z \ll \lambda_T$ in order for the system to be
effectively two-dimensional.
For small $p$, the contribution from the logarithm is suppressed
for small enough $g$, \ie by demanding
$\exp(1/g)\gg \lt / l_z$.
Neglecting the logarithmic dependence in the integral in
\pref{bu2}, one finds $\frac{d\delta_0} {dg} \approx -\frac \pi 4$.
Thus, 
in the parameter range \pref{range} there is no dependence on the 
scale $l_{z}$, and the 
Gaussian integral can be performed to get the following result for the 
leading   order in $g$ correction to the pressure,
\be{shift}
\Delta P = \frac 1 2 kT (n\lambda_T)^2 g  \ ,
\ee
in perfect agreement with \pref{eos}. We can thus conclude that for
temperatures and couplings in the range \pref{range} the scattering
approach is consistent with the mean field approximation used in the 
previous section.\footnote{This range might be experimentally relevant 
in the near future\cite{salomon}.}
It is an open question whether there are any corrections to
higher virial coefficients.

\section{Discussion}  \label{Dsec}
Since we have explicitly ignored both the possibility of a quantum
condensate and of pairing fields other than the density,
the results of this paper (and those of Ref.
\cite{bhaduri2}) can not be used for temperatures below or in the
vicinity of the Bose condensation transition $T_{c}$. This is true
irrespective of whether this transition is of the
Kosterlitz-Thouless
type or not\cite{petrov2000}.
Rather our results should be relevant in a temperature
regime where the exclusion statistics, due to the repulsive interaction,
corresponds to a small correction to the ideal Bose gas.
Moreover we have seen that one has to demand
$a \ll l_z \ll \lambda_T $ in order for the scattering approach
to be consistent with the mean field results. Since this sets a 
lower limit on the thermal wave length it also implies that the mean 
field approach can not be trusted for very rapidly changing 
potentials even at high temperatures. 

It is an interesting
open question whether the quasi particles above a
two-dimensional Bose condensate also can be described using
exclusion statistics. To answer this question one would analyze the
corresponding statistical mechanics in a more sophisticated mean
field approximation that includes effects of phase coherence and
pairing mean fields\cite{griffin}.

\vskip 2mm
\noi {\bf Acknowledgement}: S.V. thanks Prof. Christophe Salomon for a 
discussion on possible experiments.

\vspace{-0.5cm}
\bibliographystyle{unsrt}

\begin{thebibliography}{10}
\bibitem{prange1} {\em The Quantum Hall Effect}, 
                  eds. R.E. Prange and S.M. Girvin, 
                  (Second edition, Springer Verlag, New York 1990).
\bibitem{fisher1}  C. L. Kane, L. Balents and M. P. A. Fisher, 
                   Phys. Rev. Lett. {\bf 79}, 5086 (1997);  
                   R. Egger and A. Gogolin, 
                   Phys. Rev. Lett. {\bf 79}, 5082 (1997). \\
                   For a recent review, see
                   R. Egger et al. in 
                   {\sl Interacting Electrons in Nanostructures}, 
                   edited by R. Haug and H. Schoeller (Springer)
                   (cond-mat/0008008).

\bibitem{leinaas1} J.M. Leinaas and J. Myrheim,
                   Nuovo Cimento {\bf 37 B}, 1 (1977).

\bibitem{haldane1} F.D.M. Haldane,
                   Phys. Rev. Lett. {\bf 67}, 937 (1991).

\bibitem{isakov3}   S. B. Isakov,
                    Int. J. Mod. Phys. {\bf A 9}, 2563 (1994).

\bibitem{FESrealiz} Z. N. C. Ha,
                    Phys. Rev. Lett. {\bf 73}, 1574 (1994); \\
                    Z. N. C. Ha,
                    Nucl. Phys. {\bf B 435}, 604 (1995); \\
                    M. V. N. Murthy and R. Shankar,
                    Phys. Rev. Lett. {\bf 73}, 3331 (1994).
\bibitem{veigy1} A.  Dasni{\`e}res de Veigy and S.  Ouvry,
                 Phys. Rev. Lett. {\bf 72}, 600 (1994);
                 D. Li and S. Ouvry,
                 Nucl. Phys. {\bf B 430}, 563 (1994).

\bibitem{johnson1} M. D. Johnson and G. S. Canright,
                   Phys. Rev. {\bf B 49}, 2947 (1994);
                   S. He, X.-C. Xie, and F.-C. Zhang,
                   Phys. Rev. Lett. {\bf 68}, 3460 (1992).

\bibitem{hansson1}
        T. H. Hansson, J. M. Leinaas and S. Viefers,
        Nucl. Phys. {\bf B 470}, 291 (1996).

\bibitem{PRL} T. H. Hansson, J. M. Leinaas and S. Viefers,
              Phys. Rev. Lett. {\bf 86}, 2930 (2001).

\bibitem{haugset1} T. Haugset and H. Haugerud,
                Phys. Rev. {\bf A 57}, 3809 (1998).

\bibitem{assym} L. P. Pitaevskii and A. Rosch,
                Phys. Rev. {\bf A 55}, R853 (1997);
                W. J. Mullin,
                J. Low Temp. Phys. {\bf 106}, 615 (1997);
                {\it ibid} {\bf 110}, 167 (1998).


\bibitem{bhaduri2} R. K. Bhaduri, S. M. Reimann, S. Viefers, A. Ghose 
                   Choudhury and M. K. Shrivastava,
                   J. Phys. {\bf B 33}, 3895 (2000).

\bibitem{bhaduri1} R. K. Bhaduri, M. V. N. Murthy, and  M. K. Shrivastava,
                   Phys. Rev. Lett. {\bf 76}, 165 (1996).

\bibitem{isakov1} S. B. Isakov, Mod. Phys. Lett. {\bf B 8}, 319 (1994).

\bibitem{wu1} Y. S. Wu, Phys. Rev. Lett. {\bf 73}, 922 (1994).

\bibitem{isakov4} S. B. Isakov, D. P. Arovas, J. Myrheim,
                  and A. P. Polychronakos,
                  Phys. Lett. {\bf A 212}, 299 (1996).

\bibitem{nayak1} C. Nayak and F. Wilczek, 
                 Phys. Rev. Lett. {\bf 73}, 2740 (1994).

\bibitem{isakov5} S.B. Isakov, S. Mashkevich and S. Ouvry,
                  Nucl. Phys. {\bf B 448}, 457 (1995).

\bibitem{sen1} D. Sen, Nucl. Phys. {\bf B 360}, 397 (1991);
               S. Viefers, F. Ravndal, and T. Haugset,
               Am. J. Phys. {\bf 63}, 369 (1995).



\bibitem{griffin} A. Griffin, Phys. Rev. {\bf B 53}, 9341 (1996).



\bibitem{isakov2} S. B. Isakov and S. Viefers,
              Int. J. Mod. Phys. {\bf A 12}, 1895 (1997).




\bibitem{landau1} See \eg L. D. Landau and E. M. Lifshitz,
                  {\em Statistical Physics}
                  (Pergamon Press, London,1954) p. 236 for 3D. 
                  The modification to 2D is straightforward.


\bibitem{2d-delta} This is a well defined quantum problem, see \eg
                   R. Jackiw in {\it M. A. B. B\'eg Memorial Volume}
                   (World Scientific, Singapore, 1991).



\bibitem{petrov2000} D. S. Petrov, M. Holzmann, and G. V. Shlyapnikov,
                     Phys. Rev. Lett. {\bf 84}, 2551 (2000).


\bibitem{giacconi1} P. Giacconi, F. Maltoni and R. Soldati,
                                        Phys. Rev. {\bf B 53}, 10065 (1996).
                                                                                

\bibitem{salomon} C. Salomon, private communication.


\end{thebibliography}

\vspace{0.5cm}
\end{document}